\documentclass[PRB,twocolumn,superscriptaddress,showpacs]{revtex4}
\usepackage{color}
\usepackage{amssymb}
\usepackage{graphicx}

\begin{document}

\title{Temperature dependence of clusters with attracting vortices in superconducting Niobium studied by neutron scattering.}
\author{Alain Pautrat}
\affiliation{Laboratoire CRISMAT, UMR 6508 CNRS-ENSI Caen, 6 Bd Mar\'{e}chal Juin, 14050 Caen, France.}
\author{Annie Br$\hat{u}$let}
\affiliation{Laboratoire L\'{e}on Brillouin,  UMR 12 CEA-CNRS, CE Saclay, 91191 Gif sur Yvette, France.}

\begin{abstract}
We have investigated the intermediate mixed state of a superconducting niobium sample by Very Small Angle Neutron Scattering. 
  We show that this state is stabilized through a sequence where a regular vortex lattice appears, which then coexists with vortex clusters
 before vanishing at low temperature. 
 Vortices in clusters have a constant periodicity regardless of the applied field, exhibit a temperature dependence close to the one of the
 penetration depth. The clusters disappear in the high temperature limit. All the results agree with an explanation in terms of vortex
 attraction due non local effects, and indicate a negligible role of pinning.
Phase coexistence between Abrikosov vortex lattice and vortex clusters is reported showing the first order nature of the boundary line.
\end{abstract}

\newpage
\maketitle

The formation of a vortex lattice in superconductors is a subject that has always interested the scientific community. 
For superconductors of the second kind with a Ginzburg-Landau parameter $\kappa$ $>$ 1 /$\sqrt 2$, the meissner state is observed 
only below a critical field B$_{c1}$. For B $>$ B$_{c1}$ the stable state is the Shubnikov mixed phase, where only partial
 diamagnetism is observed due to the penetration of quantum vortices in the sample.
For strong enough magnetic fields, repulsive interactions stabilize a vortex lattice with long range order \cite{Abriko}.
 Physics of this Abrikosov lattice, and how it can resist the material disorder, have been heavily studied
due to both fundamental and technological issues. 
	  	
Pioneering magnetic decoration experiments have shown that the vortex state may be different in superconductors
with non zero demagnetization field and $\kappa$ close to 1/$\sqrt 2$. It was indeed observed in pure Niobium that the vortex
 lattice emerging from the surface and created by a magnetic field of a few tens of Gauss was very different from the Abrikosov
 lattice \cite{trauble}.
 Large Meissner areas coexist with clusters, where vortices
 are closer to what
 they should be in a regular mixed state. This implies that the vortex interaction has an attractive tail, 
what was found indeed possible under conditions that may
 have different origins. The case of low $\kappa$ superconductor was treated by different authors \cite{Eilenberger,Halbritter,Leung}
 and also recently discussed in \cite{Brandt} and in \cite{review}. 
In the vortex clusters, it was shown by neutron scattering that the lattice period in this intermediate mixed state remains independent
 of the magnetic field
 for a fixed temperature  \cite{Schelten1}, with a period close to 1800 $\AA$ at 4.2 K.   

 Recenly there was a growing interest in vortex states with attractive interactions.
 Vortex coalescence and attractive interaction between vortices were reported in the Spin-Triplet
 Superconductor Sr$_2$RuO$_4$ by local squid magnetometry \cite{Dolocan} and recently using muon spin rotation \cite{gray} . 
 Coexistence of vortex clusters and Meissner areas was also observed in the two gaps superconductor $MgB_2$ \cite{moshchalkov}.
 This latter observation called "Type 1.5 superconductivity" was proposed to arise from two coherence lenghts associated 
to each of the superconducting electron bands of $MgB_2$ \cite{moshchalkov}. Similar interpretation in terms of multicomponents system 
was also adressed for Sr$_2$RuO$_4$ \cite{Babaev1} and reviewed in \cite{Babaev}.
Some debate exists however on the genuine interpretation of observed topologies \cite{Brandt,comment,review}.
 As a matter of fact, the mixture of Meissner and vortex states has been reported for small magnetic fields
 of tens of Gauss, or few Gauss.
 Most of observations are made with surface studies (magnetic decoration, magneto optical imaging or local squid 
magnetometry techniques), whose experimental resolutions
 are well suited for measurements at such low fields. Magnetic decoration studies with a single vortex resolution
 have been specially efficient to produce beautiful real space pictures of the intermediate-mixed state \cite{Essman},
 but the understanding of the origin of the observed structures is not straightforward.
 Measurements are made after cooling the sample from above T$_c$ with the applied field (field cooling FC).
 First, pinning effects can not be excluded and may explain the existence of vortices for very low field  ($B<(1-D)B_{c1}$ with D
 the demagnetisant coefficient) where they are normally not expected \cite{Mumut}. 
 Under FC conditions, the observed structures are frozen at an ill-defined temperature, estimated close to T(B$_{c2}$) 
in low T$_c$'s \cite{Yanina}. This implies that the thermal variation of these structures is unknown and
that the respective role of pinning and thermodynamics is hard to quantify.
 
 One of the known candidate to explain an attractive interaction between vortices in low $\kappa$ superconductor is the non local correction.
 In a non local model,
 the potential $A(r)$ is no more simply proportional to the current density $J_s$ ($r$ is the position) \cite{Pippard}.
 One consequence is that the supercurrent efficiency is reduced compared to the pure local limit, and field penetration 
is allowed for length longer than the London penetration length $\lambda_L$.  
In a low $\kappa$ and single-gap superconductor, this non local effect may cause an attractive interaction
 between two vortices for a certain (B,T) range \cite{Kramer}. Interestingly, this non local correction should be no
 more effective close to the critical temperature
 where the London length reaches large values, and a phase boundary between the low temperature attractive region from 
the high temperature repulsive region is expected \cite{Kramer,klein}. 
Field and temperature dependent measurements are then much required and are possible with small angle neutron scattering (SANS). SANS has also the
 advantage of being a bulk probe allowing a direct measurement of the lattice periodicity and of its disorder.
The experimental resolution of SANS is however best suited for vortex lattices created by fields larger than typically 100 G. 
Neutron scattering experiments at very low fields have been then scarcely reported. 
 The intermediate state was observed at T=3.6K for magnetic fields in between 100 G and 300 G in pure Nb cylinders \cite{ted}. 
 Neutron Grating interferometry was also very recently used to visualise the morphology of heterogeneous vortex states for B$\geq $100 G
 \cite{Grunzweig}. Here, we have used the new very small angle neutron scattering (VSANS) spectrometer TPA at the Laboratoire L\'{e}on
 Brillouin (Saclay, France),
 allowing to reach low scattering vector Q such as 6.10$^{-4}$ $\AA$ $^{-1}$ in our experiments with an optimized Q resolution \cite{TPA}.
 We clarify how the vortex structures are formed at low temperature when the sample is field cooled (FC). We report on the high temperature
 boundary for attractive vortices and 
show a phase coexistence between a vortex lattice with regular periodicity and clusters of attracting vortices.

A large slab of pure Niobium (T$_c$=9.17 $\pm$ 0.05 K) was used for the neutron scatering experiment (dimensions (L=35)$\times$(w=17)$\times$(t=1.5) mm$^3$).
 A small repliqua with the same aspect ratio was cut for magnetic characterizations in a MPMS SQUID magnetometer.
 Comparison of the sample characteristics
 with litterature \cite{koch,imfeld} indicates good purity and small amount of interstitial defects 
(B$_{c2}$=2950 G, B$_{c1}$=1400 G at T=4.2K consistent with $\kappa\approx$ 0.9).
Significant critical current exists however (J$_c$ (4K, 0G)$\approx$  10$^4$ A/cm$^2$), 
largely due to the unpolished surfaces leading to important pinning \cite{surfpinning, TOF}.
  Due to the long strip shape of the sample, the first penetration field for vortices is
 largely lower than B$_{c1}$ because of a significant demagnetizing field \cite{Brandt Hp}. 
 Measurements were
 performed using the VSANS spectrometer TPA at LLB. An originality of this new spectrometer is
 its multibeam collimation composed of a large number of drilled very small holes (about $1 mm$ diameter)
 producing a small convergent beam at one point on the detector. This latter is an image plate
 of the Marresearch company, equipped with a neutrons converter; it has pixels of very small size, 0.15 $\times$ 0.15 mm$^2$, i.e. high spatial
 resolution even at low Q \cite{TPA}. Neutron wavelength was 6 $\AA$ with a Full Width at Half Maximum distribution of 14 \% . The sample to detector distance was 6.187 m. The magnetic field B
was parallel to the neutron beam and perpendicular to the large facets of the Nb sample. 
Magnetic field from 50 G to 200 G produced with a SpectroMag (10 T) superconducting magnet was applied.
 The background scattering was measured
 in the normal state (T=10 K$>T_c$) and subtracted to the raw data in order to reveal the scattering
 arising from the superconducting structures.

 A typical SANS pattern thus obtained just below $T_{c}$ at a temperature of T=8.5 K and with an applied field B=$100 G$ is shown in Fig.1.
 We observe a scattering ring characteristic of a lattice with a degraded orientational order.
 Fig.2 shows the intensity as function of the scattering vector Q at the same temperature, after a radial averaging over the 2D detector: 
a single Bragg peak can be resolved at $Q_1$. For a conventional hexagonal vortex lattice, the first 
order Bragg peak is expected at Q=2$\pi/a_{FLL}$ with $a_{FLL}=0.93 (\phi_0 / B)^{1/2}$
 for an hexagonal lattice or $a_{FLL}=(\phi_0 / B)^{1/2}$ for a square lattice ($\phi_0$
 is the quantum flux). Higher harmonics are often observed for a well ordered lattice in
 the London-like regime (typically for $\kappa^2$.B/B$_{c2}<$ 1 which is the case here) \cite{Brandt,nousBi,ted2}).  
 We do not observe a second order peak likely due to the lattice disorder.
 When the temperature decreases, at T*$\leq$  8K, a second peak emerges at $Q_2$ (Fig.3).
 Its position evolves with temperature, contrarily to $Q_1$ which is temperature independant.
  For all applied fields from 100 to 200 G, we find Q$_1$= $2\pi (B/\phi_0)^{1/2}$ as expected for a square Abrikosov lattice.
  The Q$_1$ value at 75 G is the only close to the one of the hexagonal lattice. Both square and hexagonal lattice have been already 
reported in high purity Nb close to T$_c$, 
  albeit at larger field for the square lattice \cite{ted, Ted IMS}.
 We will return to this point latter. 
  When decreasing further the temperature, the intensity of the first peak at $Q_1$ is decreasing before vanishing,
 and only the second peak at $Q_2$ subsists at low temperature as shown in Fig.4. 
We note that both peaks at $Q_1$ and $Q_2$ are observed simultaneously in a substantial temperature range (Fig.5),
 demonstrating a phase coexistence of two different lattices with different periodicities.
 We report in Fig.6 the $Q_2$ variation as function of temperature for different applied fields (B=75-200 G).
 Note that no scattered intensity (in excess to the background) was observed for B= 50 G, indicating that a full Meissner state was likely obtained. 
 All the curves in Fig.6 are
 superimposed (within the error bars at high temperature): $Q_2$ is then field independant for the whole temperature range.
 The large value of $Q_2$ implies an attractive interaction between the flux lines.
The corresponding periodicity, $d_2=2\pi/Q_2$, can be interpreted as the upper limit for the lattice spacing as expected
 in this regime \cite{brandt}. A similar conclusion was reported from the analysis of the induction jump at B$_{c1}$ in (pinning free)
 low $\kappa$ superconductor with highly reversible magnetization \cite{Dichtel,Ulmaier,Finnemore,Weber} or from neutron diffraction experiment
 for temperature close to 4 K \cite{Schelten1, Ted IMS}. Fig.7 shows the corresponding lattice periodicity 
 $d_2$ as function of the temperature, independently of the applied field up to 200 G. The value $d_2$= 1660 $\AA$ measured at 4 K is in good agreement 
with the values reported by Schelten et al. ($\approx$ 1800 $\AA$) and M$\ddot{u}$hlbauer et al. ($\approx$  1600 $\AA$)
 \cite{Schelten1, Ted IMS}. 
 Theoretical calculations predict that the temperature dependence of the equilibrium distance between attracting vortices
 is governed by the London penetration depth \cite{Dichtel,Brandt}. We have tried the Casimir-Gorter variation
 $y=(1-t^4)^{-1/2}$ with $t=T/T_c$  \cite{Finnemore2}, and an analytic form $y=(1-t^{(3-t)})^{-1/2}$ which follows closely the BCS variation \cite{BCS}.
 As shown in the Fig.7, the latter expression corresponds well to the thermal variation of $d_2$ up to the temperature $T^{*}$ where the $Q_2$ peak vanishes. 
In our experiments, the temperature interval between the recorded data is relatively large and $T^{*}$ is not very precisely determined.
 However, we find $T^{*}/T_c \approx$ 0.87-0.92, in reasonable agreement with the boundary line between the dominating
 vortex attraction and the dominating vortex repulsion regimes
 predicted by non local models \cite{Kramer, klein, Ulmaier}. The phase coexistence observed here is a bulk probe of the first order
 nature of the boundary line between the two regimes, and is in agreement with interpretations of the magnetization
 discontinuity in superconductors with small $\kappa$ parameters \cite{Ulmaier}. Since we observe that the attractive interaction disapears at high temperature,
 it can not be the mechanism which stabilizes the square lattice at low field in our sample. The effect of fourfold Fermi surface
 symmetry can be already relevant \cite{nakai}. Considering the differences in the lattice symetry at low field and high temperature
 reported here and in \cite{Ted IMS},
 it confirms the close competition between the cubic crystal symmetry and the tendency of vortex
 to form hexagonal lattice existing in Niobium and that small differences of quality between samples apparently changes the dominant interaction. 

Analysis of the $Q_2$ width shows a peak broadening much above the experimental resolution and then a non perfect positional ordering.
To extract a quantitative value of the positional correlation length $\tau_2$, the instrumental resolution has to be accounted for.
The Q resolution of the non standard multibeam VSANS instrument TPA and the calculation of $\tau_2$ are given in the supplementary material. 
The correlation length $\tau_2$ is found to be close to 1.2 $\mu$m at low temperature (see Fig.8). It is similar to the typical clusters size
 reported from magnetic decorations experiments \cite{Essman}. We can then conclude that the bulk structures are
 not strongly different from the structures observed at the surfaces, and at least that the Landau branching of vortex domains at the surface is small.
 We observe that $\tau_2$ is roughly constant only below $T\leq T_c/2$, a temperature which is the temperature 
below which there is no significant changes of the condensate parameters, such as the penetration depth. Thus, no pinning effects 
are required to explain the quasi-constant  clusters size of attractive vortices at low temperature.

 Since our sample has a significant critical current, we have to discuss if bulk pinning 
may contribute to disordering of the lattice and to the existence of vortex clusters,
notwithstanding the attractive interaction. The dominant pinning mechanism in pure Nb is generally surface pinning \cite{surfpinning, TOF} 
which has no direct relation with a bulk disordering of the lattice \cite{noussans}. As a consequence, significant critical current can be
 present without direct relation with vortex disorder due to bulk pinning. This is the situation which makes most sense here considering the
 good bulk purity of our sample which has superconducting parameters very close to the intrinsic values of Nb.
 We have also shown in this experiment that the clusters have characteristics that closely follow equilibrium theories.
 All this indicates that pinning only plays a minor role for stabilizing the vortex structures observed here.

In conclusion, we have measured the temperature dependence of clusters formed by vortices  
 under attractive interaction by neutrons scattering. This interaction is no more effective close to $T_c$, showing 
 that it originates from non local effects. The phase coexistence between Abrikosov lattice and vortex clusters demonstrates
 the first order nature of the transition. 
Since the temperature dependence of vortex periodicity in these clusters follows the one of the penetration depth,
 pinning does not play the leading role in the stabilization of the vortex cluster structures observed at low temperature and low field.

\newpage
\vskip 2 cm

\begin{figure}[htbp]
\includegraphics*[width=5cm]{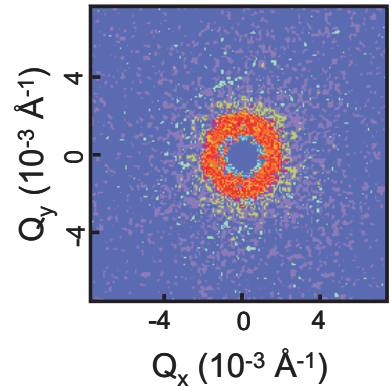}
\caption{2D VSANS pattern of a pure Niobium sample obtained at T= 8.5 K, after a field cooling at B=100 G.
 The 2D pattern obtained at high temperature (T=10 K) has been subtracted. The scattering ring arises from the disordered flux line lattice.}
\end{figure}

\begin{figure}[htbp]
\includegraphics*[width=7cm]{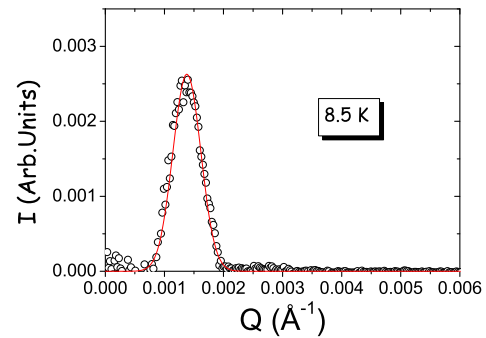}
\caption{Scattered intensity as function of the scattering vector Q at T=8.5 K, after a 100 G field cooling. One Bragg peak is observed at Q$_1$=1.38 10$^{-3}\AA^{-1}$}
\end{figure}

\begin{figure}[htbp]
\includegraphics*[width=7cm]{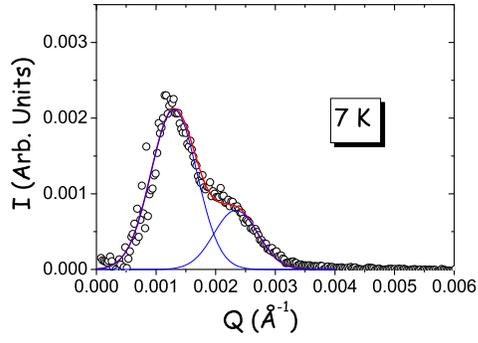}
\caption{Scattered intensity as function of the scattering vector Q at T=7 K, after a 100 G field cooling. Two peaks are observed at Q$_1$=1.36 10$^{-3}\AA^{-1}$ and Q$_2$=2.34 10$^{-3}\AA ^{-1}$.}
\end{figure}

\begin{figure}[htbp]
\includegraphics*[width=7cm]{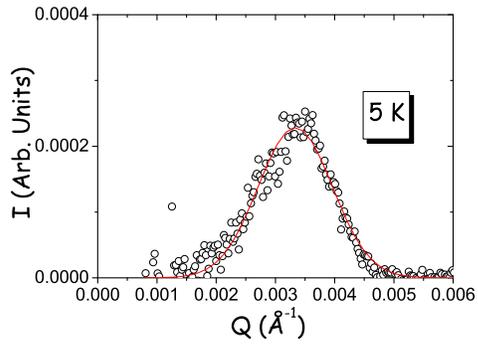}
\caption{Scattered intensity as function of the scattering vector Q at T=5 K, after a 100 G field cooling.
A single peak is observed at Q$_2$=3.34 10$^{-3}\AA ^{-1}$.}
\end{figure}

\begin{figure}[htbp]
\includegraphics*[width=7cm]{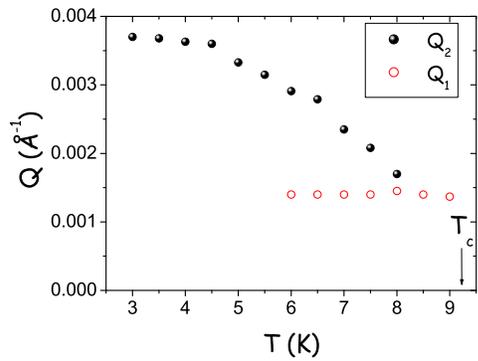}
\caption{Variation of Q$_1$ and Q$_2$ as function of the temperature (after a 100 G FC).}
\end{figure}

\begin{figure}[htbp]
\includegraphics*[width=7cm]{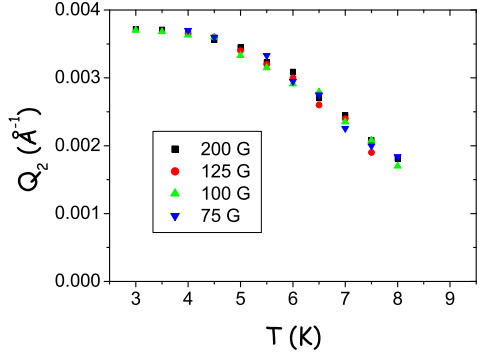}
\caption{Variation of Q$_2$ as function of the temperature for different applied magnetic field (B=75, 100, 125 and 200 G).}
\end{figure}

\begin{figure}[htbp]
\includegraphics*[width=7cm]{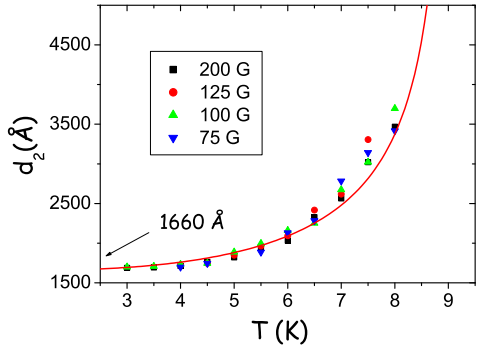}
\caption{Variation of d$_2$, the lattice spacing in the vortex clusters, as function of the temperature for different
 applied magnetic field (B=75, 100, 125 and 200 G). The line corresponds to the BCS-like dependence.}
\end{figure}

\begin{figure}[htbp]
\includegraphics*[width=7cm]{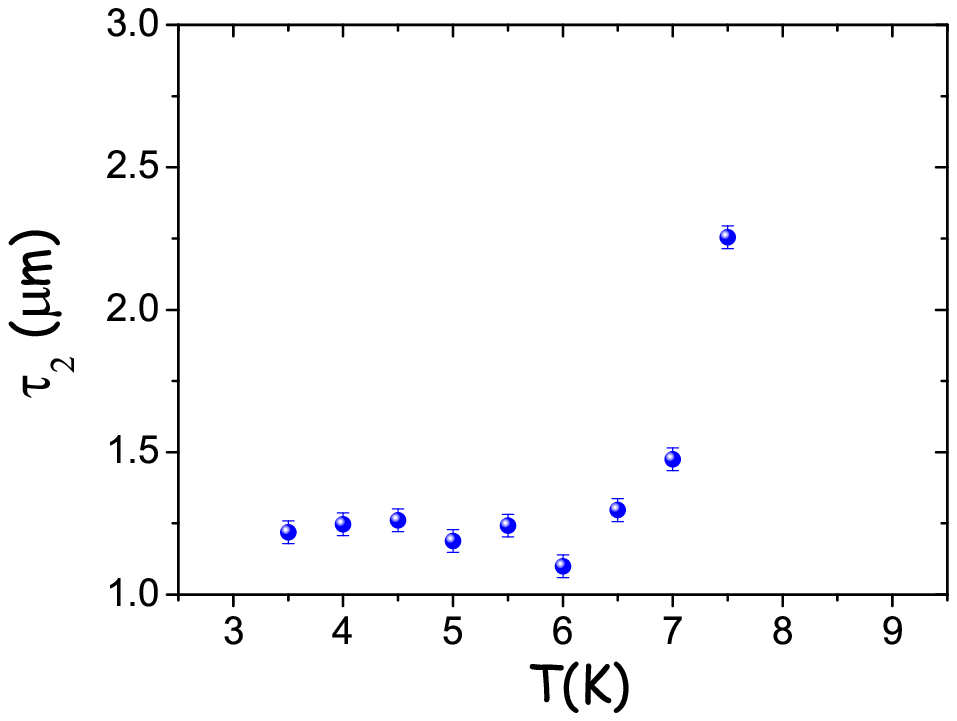}
\caption{Variation of $\tau_2$, the typical size of clusters composed of vortices with attractive interactions, as function of the temperature.}
\end{figure}


\begin{references}

\label{sec:TeXbooks}
\bibitem{Abriko} A.A. Abrikosov, Soviet Physics JETP 5, 1174 (1957).
\bibitem{trauble} H. Tr$\ddot{a}$uble and U. Essmann, Phys. Stat. Sol. 20, 95 (1967).
                  U. Kr$\ddot{a}$geloh, Phys. Lett. A 28, 657 (1969).
\bibitem{Eilenberger} G. Eilenberger and H. Buttner, Z. Physik 224, 335 (1969).
\bibitem{Halbritter} J. Halbritter, Z. Physik 243, 201 (1971).
\bibitem{Leung} M.C. Leung, Journal of Low Temp. Phys. 12, 215 (1973).
\bibitem{Brandt} E. H. Brandt and M. P. Das, Journal of Superconductivity and Novel Magnetism 24, 57 (2011). 
\bibitem{review} E. Babaev and M. Silaev, J. Supercond Nov Magn 26, 2045 (2013).
\bibitem{Schelten1} J. Schelten, U. Ullmaier and W. Schmatz, Phys. Status Solidi (B) 48, 619 (1971).
\bibitem{Dolocan} V. O. Dolocan, C. Veauvy, F. Servant, P. Lejay, K. Hasselbach, Y. Liu, and D. Mailly, Phys. Rev. Lett. 95, 097004 (2005).
\bibitem{gray} S. J. Ray, A. S. Gibbs, S. J. Bending, P. J. Curran, E. Babaev, C. Baines, A. P. Mackenzie, and S. L. Lee, Phys. Rev. B 89, 094504 (2014).
\bibitem{moshchalkov} V. Moshchalkov, M. Menghini, T. Nishio, Q. H. Chen, A. V. Silhanek, V. H. Dao, L. F. Chibotaru,
 N. D. Zhi-gadlo, and J. Karpinski, Phys. Rev. Lett. 102, 117001 (2009).
\bibitem{Babaev1} J. Garaud, D. F. Agterberg, and E. Babaev, Phys. Rev. B 86, 060513(R) (2012).
\bibitem{Babaev} E. Babaev, J. Carlstrom, J. Garaud, M. Silaev, J.M. Speight, Physica C: Superconductivity 479, 2 (2012).                 
\bibitem{comment} E. Babaev and M. Silaev, Phys. Rev. B 86, 016501 (2012); V. G. Kogan and J. Schmalian, ibid. 86, 016502 (2012). 
\bibitem{Essman} U. Krageloh, Phys. Lett. 28 A, 657 (1969); 
                 U. Essmann, Phys. Lett. 41A, 477 (1972).                 
\bibitem{Mumut} E. H. Brandt and S.-P. Zhou, Physics 2, 22 (2009).
\bibitem{Yanina} Y. Fasano and M. Menghini, Supercond. Sci. Technol. 21, 023001 (2008).
\bibitem{Pippard}  A.B. Pippard, Proc. R. Soc. Lond. A 24, 547 (1953). 
\bibitem{Kramer} L. Kramer, Z. Phys. 258, 367 (1973); M. C. Leung and A. E. Jacobs, in Low Temperature Physics LT-13 (Plenum, New York, 1974), p. 46.
\bibitem{klein}  U. Klein, J. Low Temp. Phys. 69, 1 (1987). 
\bibitem{ted} M. Laver, C. J. Bowell, E. M. Forgan, A. B. Abrahamsen, D. Fort, C. D. Dewhurst, S. M$\ddot{u}$hlbauer, D. K. Christen, J. Kohlbrecher, R. Cubitt, and S. Ramos, Phys. Rev. B 79, 014518 (2009).
\bibitem{Grunzweig} C. Gr$\ddot{u}$nzweig, S. M$\ddot{u}$hlbauer, M. Schulz, M. Gr$\ddot{u}$nzweig, A. Kaestner, J. Kohlbrecher, P. Mikheenko, T. Reimann, U. Keiderling, P. Boeni, arXiv:1308.3612 (2013). 
\bibitem{TPA} A. Br$\hat{u}$let, V. Thevenot, D. Lairez, S. Lecommandoux, W. Agut, S.P. Armes, J. Duc, S. Desert, J. Appl. Cryst. 41, 161 (2008). 
\bibitem{koch} C. C. Koch, J. O. Scarbrough et D. M. Kroeger, Phys. Rev. B 9 888 (1974).
\bibitem{imfeld} N. J. Imfeld, W. Bestgen, and L. Rinderer, Journal of Low Temperature Physics 60, 223 (1985). 
\bibitem{surfpinning} A. Pautrat, J. Scola, C. Goupil, Ch. Simon, C. Villard, B. Domenges, Y. Simon, C. Guilpin, and L. Mechin,
 Phys. Rev. B 69, 224504 (2004).
\bibitem{TOF} Alain Pautrat, Annie Brulet, Charles Simon, and Patrice Mathieu, Phys. Rev. B 85 (2012) 184504. 
\bibitem{Brandt Hp} E. H. Brandt,Phys.Rev.B59, 3369 (1999).
\bibitem{nousBi} A. Pautrat, Ch. Simon, C. Goupil, P. Mathieu, A. Br$\hat{u}$let, C. D. Dewhurst, and A.I. Rykov, Phys. Rev. B 75, 224512 (2007).
\bibitem{ted2} E. M. Forgan, S. J. Levett, P. G. Kealey, R. Cubitt, C. D. Dewhurst, and D. Fort, Phys. Rev. Lett. 88, 167003 (2002).
\bibitem{Ted IMS} S. M$\ddot{u}$hlbauer, C. Pfleiderer, P. Buni, M. Laver, E. M. Forgan, D. Fort, U. Keiderling, and G. Behr, Phys. Rev. Lett.
 102, 136408 (2009).
\bibitem{brandt} E. H. Brandt, Phys. Rev. B 18, 6022 (1978); Phys. Rev. Lett. 78, 2208 (1997).
\bibitem{Dichtel} K. Dichtel, Phys. Lett. 35A, 285 (1971).
\bibitem{Ulmaier} J. Auer and H. Ullmaier, Phys. Rev. B 7, 136 (1973).
\bibitem{Finnemore} D.K. Finnemore, J.R. Clem and T.F. Stromberg, Phys. Rev. B 6, 1056 (1972).
\bibitem{Weber} H.W. Weber et al, Phys. C 161, 272 (1989).
\bibitem{Finnemore2} D.K. Finnemore, T.F. Stromberg, and C.A. Swenson, Phys. Rev. 149, 231 (1966).
\bibitem{BCS} B. Muhlschlegel, Z. Phys. 155, 313 (1959)
\bibitem{nakai} N. Nakai, P. Miranovic, M. Ichioka, and K. Machida, Phys. Rev. Lett. 89, 237004 (2002).
\bibitem{noussans}A. Pautrat, M. Aburas, Ch. Simon, P. Mathieu, A. Br$\hat{u}$let, C. D. Dewhurst, S. Bhattacharya, and M. J. Higgins, Phys. Rev. B 79, 184511 (2009);
 A. Pautrat, J. Scola, C. Simon, P. Mathieu, A. Br$\hat{u}$let, C. Goupil, M. J. Higgins, and S. Bhattacharya, ibid. 71, 064517 (2005).
\end{references}
\end{document}